# Advancing Deformable Medical Image Registration with Multi-axis Cross-covariance Attention

Mingyuan Meng [a, b], Michael Fulham [a, c], Lei Bi [a, b, *], and Jinman Kim [a, *]

[a] School of Computer Science, The University of Sydney, Australia.
[b] Institute of Translational Medicine, Shanghai Jiao Tong University, Shanghai, China.
[c] Department of Molecular Imaging, Royal Prince Alfred Hospital, Australia.
[*] Corresponding author: lei.bi@sjtu.edu.cn; jinman.kim@sydney.edu.au

**Abstract** — Deformable image registration, aiming to find a dense (pixel-wise) non-linear spatial correspondence between images, is a fundamental requirement for medical image analysis. Recently, transformers have been widely used in deep learning-based registration methods for their ability to capture long-range dependency via self-attention (SA). However, the high computation and memory loads of SA (growing quadratically with the spatial resolution) hinder transformers from processing subtle textural information in high-resolution image features, e.g., at the full and half image resolutions. This limits deformable registration as the high-resolution textural information is crucial for finding precise pixel-wise correspondence between subtle anatomical structures. Cross-covariance Attention (XCA), as a "transposed" version of SA that operates across feature channels, has complexity growing linearly with the spatial resolution, providing the feasibility of capturing long-range dependency among high-resolution image features. However, existing XCA-based transformers merely capture coarse global long-range dependency, which are unsuitable for deformable image registration relying primarily on fine-grained local correspondence. In this study, we propose to improve existing deep learning-based registration methods by embedding a new XCA mechanism. To this end, we design an XCA-based transformer block optimized for deformable medical image registration, named Multi-Axis XCA (MAXCA). Our MAXCA serves as a general network block that can be embedded into various registration network architectures. It can capture both global and local long-range dependency among high-resolution image features by applying regional and dilated XCA in parallel via a multi-axis design. Extensive experiments on two well-benchmarked inter-/intra-patient registration tasks with seven public medical datasets demonstrate that our MAXCA block enables state-of-the-art registration performance.

## 1. Introduction

Medical image registration is a fundamental step for medical image analysis and has been an active research focus for decades (Sotiras, Davatzikos, and Paragios 2013; Zou et al. 2022). It spatially aligns medical images acquired from different patients, times, or scanners, which serves as a crucial step for various clinical tasks, including tumor growth monitoring and group analysis (Haskins, Kruger, and Yan 2020). Deformable image registration aims to find a dense (pixel-wise) non-linear spatial transformation between a pair of images, through which the two images can be spatially aligned after warping. Traditional methods typically formulate deformable image registration as a time-consuming iterative optimization problem (Avants et al. 2008; Modat et al. 2010). Recently, deep registration methods based on Convolutional Neural Networks (CNNs) and/or transformers have been widely recognized for fast end-to-end registration (Haskins, Kruger, and Yan 2020; Zou et al. 2022). These methods learn a mapping from image pairs to spatial transformations based on a set of training data, which have shown superior registration performance than traditional registration methods (Zou et al. 2022).

Visual transformer (Dosovitskiy et al. 2021) and its window-based variant, Swin transformer (Liu et al. 2021), have been widely used in vision tasks for the capability to capture long-range dependency via self-attention (SA). This capability enables transformers to surpass CNNs in deformable image registration as it enables larger receptive fields to capture large deformations between images (Chen et al. 2022; Chen, Zheng, and Gee 2023; Ma et al. 2023; Wang, Ni, and Wang 2023; Zhu and Lu 2022). However, the high computation and memory loads of SA (growing quadratically with the spatial resolution) hinder transformers from processing subtle textural information that is available only in high-resolution image features, e.g., at the full and half image resolutions (Meng et al. 2023b). This limitation is non-negligible for deformable medical image registration, as the high-resolution textural information is crucial for identifying subtle anatomy in medical images and finding precise pixel-wise spatial correspondence between anatomical structures (Meng et al. 2024). To compensate for this limitation, recent hybrid `registration methods employed convolutional layers to process high-resolution image features while using transformers at downsampled resolutions after patch embedding (Chen et al. 2022; Chen, Zheng, and Gee 2023; Ma et al. 2022). Unfortunately, convolutional layers have difficulties in capturing long-range dependency due to the intrinsic locality of convolution operations and the lack of global connectivity (Li et al. 2021).

To address this limitation, Multi-layer Perceptron (MLP)-based registration methods were recently proposed and attained better performance than existing transformer-based registration methods (Meng et al. 2023b; Meng et al. 2024). These methods used MLP blocks at the full image resolution to capture fine-grained long-range dependency, providing long-range contextual information with enriched anatomical details. To this end, MLP blocks discard attention mechanisms to reduce the computational complexity; however, the benefits of attention mechanisms have been widely recognized in many medical vision tasks including deformable registration (Chen et al. 2023; Li et al. 2023). Moreover, to further reduce the computation, MLP blocks calculate spatial interactions by applying MLPs in a channel-wise manner (Tolstikhin et al. 2021; Tu et al. 2022), which may limit the modeling of spatial interactions to each single feature channel. These drawbacks drive us to rethink: Is there an alternative solution that allows the modeling of high-resolution long-range dependency via attention mechanisms?

Cross-covariance Attention (XCA) is a "transposed" version of SA that operates across feature channels rather than tokens, where the interactions are based on the cross-covariance matrix between keys and queries (Ali et al. 2021). As its computational complexity grows linearly with the spatial resolution, XCA provides the feasibility of capturing long-range dependency among high-resolution image features via attention mechanisms. For example, XCA-based transformers have been used to capture global long-range dependency at the full image resolution in natural image restoration tasks (Zamir et al. 2022). Unfortunately, XCA has not been explored and optimized for image registration tasks: existing XCA-based transformers merely capture coarse global long-range dependency, which are unsuitable for deformable medical image registration relying primarily on fine-grained local correspondence.

In this study, we propose a new XCA-based transformer block optimized for deformable medical image registration, named Multi-Axis XCA (MAXCA), which improves existing deep registration methods by embedding a new XCA mechanism. Our MAXCA serves as a general network block that can be embedded into various registration network architectures to capture both global and local long-range dependency among high-resolution image features. To achieve this, this block splits the input image feature maps into local regions and then captures long-range dependency by applying regional/dilated XCA within/across these regions in two parallel branches with a multi-axis design. Our main contributions are summarized as follows:

• We investigate the optimized use of XCA for deformable medical image registration, to the best of our knowledge, which is the first investigation revealing the potential of XCA for image registration tasks.

• We propose MAXCA, the first XCA-based transformer block that is specifically optimized for deformable medical image registration tasks to capture both global and local long-range dependency among high-resolution image features.

• By embedding MAXCA blocks into various registration network architectures, we develop the first set of XCA-based registration networks and attain consistent improvements over their CNN/SA-based counterparts.

Extensive experiments were conducted on two well-benchmarked medical registration tasks (3D inter-patient brain image registration and 4D intra-patient cardiac image registration) with seven public medical datasets, which demonstrated that our MAXCA block produced consistent improvements in various registration network architectures and achieved superior performance over state-of-the-art deformable medical image registration methods.

## 2. Related Works

### 2.1. Deformable Medical Image Registration

In the era of deep learning, deep registration methods commonly use convolutional layers or transformers as the basic unit to build registration networks. VoxelMorph (Balakrishnan et al. 2019), as one of the most commonly benchmarked deep registration methods, used a hierarchical encoder-decoder CNN similar to Unet (Çiçek et al. 2016), motivating the wide use of CNNs and encoder-decoder architectures in subsequent studies (Dalca et al. 2019; Jia et al. 2022; Meng et al. 2022b; Mok and Chung 2020a).

In recent years, transformers have been widely used for deformable medical image registration. For example, Swin-VoxelMorph (Zhu and Lu 2022) used a pure transformer-based registration network similar to Swin-Unet (Cao et al. 2022). TransMorph (Chen et al. 2022) used a hybrid CNN-transformer registration network, where Swin transformers were employed after 4×4×4 patch embedding and convolutional layers were employed at the full and half image resolutions. These methods used transformers at downsampled resolutions to reduce the computation/memory loads. An exception is ModeT (Wang, Ni, and Wang 2023), where motion decomposition transformers were applied at the full image resolution by computing SA operations within the local neighborhood (kernel size=3) around each pixel. However, this approach inevitably compromised the transformer's capability to capture long-range dependency.

Recently, MLPs were introduced for deformable registration as superior alternatives to transformers. MLPMorph (Meng et al. 2023b) used the same encoder-decoder architecture as VoxelMorph/TransMorph but employed MLPs in the encoder to capture fine-grained long-range dependency beginning from the full resolution. This enabled MLP-Morph to gain better performance than its CNN-/transformer-based counterparts (VoxelMorph/TransMorph).

Network architecture is also a crucial factor influencing registration performance. Many deep registration methods used the basic Unet-like direct registration architecture, e.g., VoxelMorph. Moreover, progressive registration architecture has also been widely used in recent coarse-to-fine deep registration methods to handle large deformations. For example, LapIRN (Mok and Chung 2020b) cascaded multiple laplacian pyramid networks to perform multiple registration steps. NICE-Net (Meng et al. 2022a) used a pyramid network to perform coarse-to-fine registration in a single network iteration, and it has been extended to a transformer-based variant, NICE-Trans (Meng et al. 2023a). There also exist more complicated architecture designs. For instance, CorrMLP (Meng et al. 2024) used MLP blocks in a correlation-aware coarse-to-fine architecture, which incorporated correlation information into progressive registration and attained state-of-the-art performance.

### 2.2. Cross-covariance Attention (XCA)

The concept of XCA was proposed in the cross-covariance image transformer (XCiT) (Ali et al. 2021), which achieved competitive performance with conventional SA-based transformers on natural image classification, detection, and segmentation. Subsequently, XCA was optimized for natural image restoration tasks (e.g., image denoising and deblurring) in an efficient transformer model, named Restormer (Zamir et al. 2022). The Restormer used XCA-based transformer blocks to capture global long-range dependency among high-resolution image features, which enabled it to achieve superior performance over SA-based transformer models. Recently, XCA was also used for medical image classification, where a residual XCA-based transformer was proposed to extract spatial and global features from ultrasound images (Sarker et al. 2023). These XCA-based transformer models calculated XCA on the entire image space, which merely captured coarse global long-range dependency and, unfortunately, are unsuitable for deformable medical image registration.

## 3. Method

Deformable image registration aims to find a spatial transformation $\phi$ that warps a moving image $I_m$ to a fixed image $I_f$, so that the warped image $I_{m \circ \phi} = I_m \circ \phi$ is spatially aligned with the $I_f$. In this study, we assume the $I_m$ and $I_f$ are two single-channel, grayscale volumes defined in a 3D spatial domain $\Omega \subset \mathbb{R}^3$. The $\phi$ is parameterized as a displacement field, and we parametrized the image registration problem as a function $\mathcal{R}_\theta(I_f, I_m) = \phi$, with $\theta$ as a set of learnable parameters. The function $\mathcal{R}_\theta$ is implemented as an XCA-based registration network built upon our Multi-Axis XCA (MAXCA) blocks (refer to Section 3.1). Our MAXCA block is readily embedded into various network architectures, and we exemplified it in three different architectures (refer to Section 3.2). The learnable parameters $\theta$ are optimized through unsupervised learning (refer to Section 3.3).

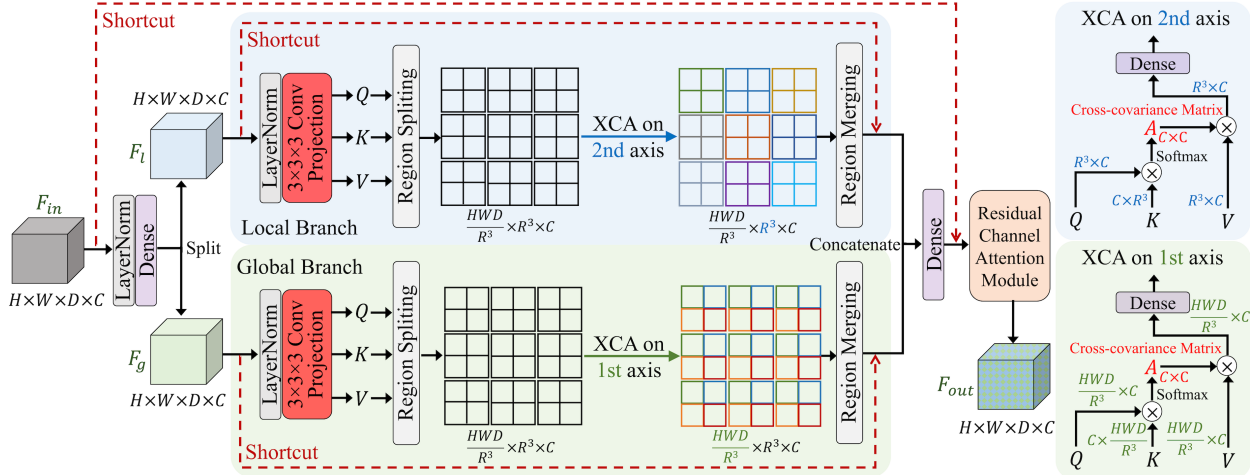

Figure 1. Illustration of Multi-Axis XCA (MAXCA) block. It first splits the input image feature maps into local regions and then applies XCA on different axes of the regional features to capture global and local long-range dependency.

### 3.1. Multi-Axis XCA (MAXCA) block

Our MAXCA block is inspired by the multi-axis blocked self-attention (Zhao et al. 2021) and multi-axis gated MLP (Tu et al. 2022). These methods performed SA/MLP on two axes to realize two forms of sparse operations, namely regional and dilated SA/MLP, to capture local and global information. However, these SA/MLP-based methods were not designed for registration tasks, and they are inherently limited when applied to deformable medical image registration (as discussed in Section I). Further, the multi-axis design has also not been investigated in the context of XCA. Here, we optimize XCA with the concept of "multi-axis" for deformable image registration by developing a Multi-Axis XCA (MAXCA) block.

The MAXCA block is illustrated in Figure 1. We assume that the input feature map $F_{in}$ has the size of $H \times W \times D \times C$. The $F$ is first projected to increase its channel number to $2C$ and then diverges into two feature heads, $F_l$ and $F_g$, in two parallel local and global branches. In the two branches, the feature maps are split into local regions according to a given region size $R$, resulting in regional feature maps in size of $(HWD/R^3) \times R^3 \times C$ with non-overlapping regions each with size of $R^3$. In the local branch, XCA is calculated on the 2nd axis (i.e., regional XCA within the regions); in the global branch, XCA is calculated on the 1st axis (i.e., dilated XCA across the regions). Intuitively, applying XCA in the two parallel branches corresponds to local and global attention in the feature map. Moreover, we propose to use convolutional QKV projection, in place of the common linear projection, to enhance the local context before computing feature covariance. To this end, we use 3×3×3 convolutional layers to project the $F_l$ and $F_g$ into $Q$, $K$, and $V$ before the region splitting. After the XCA calculation, the processed features are merged to restore the original shape.

The processed features derived from the local and global branches are concatenated and then are projected to reduce the channel number to $C$, with a residual connection from $F_{in}$. In addition, we followed the previous study (Meng et al. 2024) to apply a residual channel attention module to highlight crucial feature channels, consisting of layer normalization, convolutional layers, LeakyReLU activation, and squeeze-and-excitation (SE) channel-wise attention (Hu, Shen, and Sun 2018), with a residual connection.

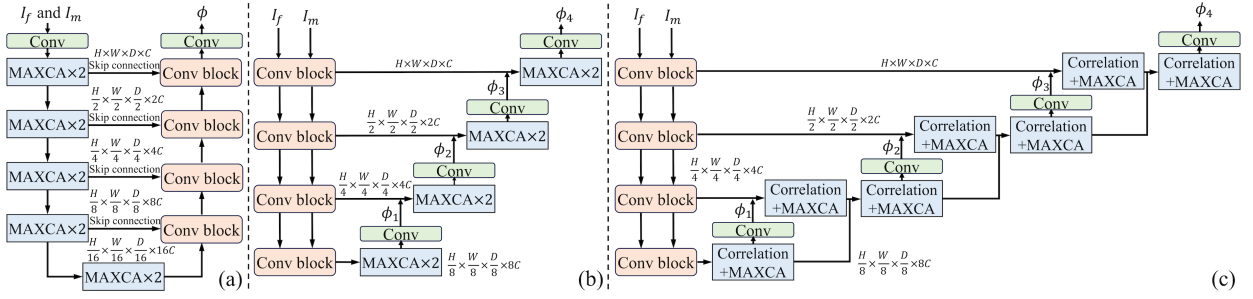

Figure 2. Illustration of three exemplified registration network architectures: (a) Direct registration (XCAMorph). (b) Progressive registration (NICE-XCA). (c) Correlation-aware progressive registration (CorrXCA).

Compared to the widely used SA with quadratic computational complexity $O(H^2W^2D^2)$, XCA has linear complexity as it relies on the cross-covariance matrix to model the feature interactions (Ali et al. 2021). The cross-covariance matrix operates across feature channels and thus has a much smaller size than the attention map calculated by SA, i.e., $C \times C$ vs $HWD \times HWD$. Existing XCA-based transformers applied the small cross-covariance matrix on the entire image space, only capturing coarse global dependency without sufficient capability to model the interactions among subtle local details. In contrast, our MAXCA block applies a cross-covariance matrix (with the size of $C \times C$) to each local region, enforcing it to model the local interactions within the region. Meanwhile, the global interaction across the regions can be captured by the global branch. With this design, our MAXCA can leverage the covariance matrix to model subtle local correspondence while not sacrificing the capability to capture global long-range interactions, which is crucial for deformable medical image registration.

### 3.2. Network Architecture

Our MAXCA block is exemplified in three network architectures: direct, progressive, and correlation-aware progressive registration architectures, as illustrated in Figure 2. Detailed architecture settings, including feature dimensions, head numbers, and region size, are provided in Appendix A.

We adopted the direct registration architecture widely used in existing deep registration methods including VoxelMorph (Balakrishnan et al. 2019), TransMorph (Chen et al. 2022), and MLPMorph (Meng et al. 2023b). It employs an Unet-style encoder-decoder network to realize a direct mapping from the input images $I_f/I_m$ to the displacement field $\phi$. As shown in Figure 2(a), we employed MACXA blocks at the encoder and denote this network as XCAMorph, following VoxelMorph, TransMorph, and MLPMorph. A 3×3×3 convolutional layer is used before the first MAXCA block to convert the input images into initial feature maps. Patch merging modules are used at the encoder to downsample the feature maps between two MAXCA blocks. The decoder is composed of successive Conv blocks and upsampling operations. Each Conv block contains two 3×3×3 convolutional layers followed by LeakyReLU activation with a parameter of 0.2 and instance normalization.

We adopted the progressive registration architecture used in recent coarse-to-fine deep registration methods, NICE-Net (Meng et al. 2022a) and NICE-Trans (Meng et al. 2023a). It consists of a CNN-based encoder that extracts two feature pyramids from $I_f/I_m$ and a progressive registration decoder that performs multiple steps of coarse-to-fine registration (refer to NICE-Trans for detailed descriptions). As shown in Figure 2(b), we employed MAXCA blocks at the progressive registration decoder and denote this network as NICE-XCA, following NICE-Net and NICE-Trans. For comparison, we also build an MLP-based progressive registration network (denoted by NICE-MLP) with Swin-MLP blocks (Liu et al. 2021) employed at the progressive registration decoder.

We also adopted the correlation-aware progressive architecture that was recently proposed in CorrMLP (Meng et al. 2024). It consists of a CNN-based encoder and a correlation-aware progressive registration decoder, which incorporates image-level and step-level correlation information into the progressive registration architecture (refer to CorrMLP for detailed descriptions). As shown in Figure 2(c), we employed MAXCA blocks at the correlation-aware progressive registration decoder and denote this network as CorrXCA, following its MLP-based predecessor, CorrMLP.

### 3.3. Unsupervised Training

To remove the reliance on ground truth labels, recent deep registration methods tend to use image similarity metrics as the training loss to perform fully unsupervised training. We followed the common unsupervised training scheme for a fair comparison: The learnable parameters $\theta$ are optimized using an unsupervised loss $\mathcal{L}$ that does not require labels. The $\mathcal{L}$ is defined as $\mathcal{L} = \mathcal{L}_{sim} + \sigma \mathcal{L}_{reg}$, where the $\mathcal{L}_{sim}$ is an image similarity term that penalizes the differences between the warped image $I_{m \circ \phi}$ and the fixed image $I_f$, the $\mathcal{L}_{reg}$ is a regularization term that encourages smooth and invertible transformations $\phi$, and the $\sigma$ is a regularization parameter that is set as 1 by default.

We adopted negative local normalized cross-correlation (NCC) as the $\mathcal{L}_{sim}$, which is a commonly used similarity metric in deformable registration methods (Balakrishnan et al. 2019; Chen et al. 2022; Jia et al. 2022; Meng et al. 2024). For the $\mathcal{L}_{reg}$, we imposed a diffusion regularizer on the $\phi$ to encourage its smoothness. We adopted these common loss functions for a fair comparison with existing deep registration methods, while other loss functions can also be easily embedded in our method to enable, e.g., diffeomorphic registration or multi-modal image registration.

## 4. Experimental Setup

### 4.1. Datasets and Preprocessing

Our method was evaluated on inter-patient brain image registration and intra-patient cardiac image registration, involving seven public medical image datasets:

For brain image registration, we adopted six public 3D brain MRI datasets that have been widely used to evaluate medical image registration (Meng et al. 2024). A total of 2,656 brain MRI images acquired from four public datasets, ADNI (Mueller et al. 2005), ABIDE (Di Martino et al. 2014), ADHD (ADHD-200 consortium 2012), and IXI (IXI dataset 2022), were used for training; two public brain MRI datasets with anatomical segmentation, Mindboggle (Klein and Tourville 2012) and Buckner (Fischl 2012), were adopted for validation and testing. The Mindboggle dataset contains 100 MRI images and was randomly split for validation/testing with a ratio of 50%/50%. The Buckner dataset contains 40 MRI images and was used for independent testing. We performed inter-patient registration for evaluation, where 100 image pairs were randomly picked from each of the Mindboggle and Buckner testing sets, resulting in 200 testing image pairs in total. We performed standard brain MRI preprocessing procedures, including brain extraction, intensity normalization, and affine registration by FreeSurfer (Fischl 2012) and FLIRT (Jenkinson and Smith 2001). All images were affine-transformed and resampled to align with the MNI-152 brain template (Fonov et al. 2011) with 1mm isotropic voxels, and then were cropped into 144×192×160.

For cardiac image registration, we adopted the public ACDC dataset (Bernard et al. 2018) that contains 4D cardiac cine-MRI images of 150 patients. Each 4D cine-MRI image contains tens of 3D frames acquired from different time points, including End-Diastole (ED) and End-Systole (ES) frames with segmentation labels of the left ventricular cavity, right ventricular cavity, and myocardium. The ACDC dataset provides 100 cine-MRI images in the training set and 50 cine-MRI images in the testing set. We randomly divided the training set into 90 and 10 cine-MRI images for training and validation and used the provided testing set for testing. The intra-patient ED and ES frames were registered with each other (ED-to-ES and ES-to-ED), resulting in 100 testing image pairs derived from the testing set. All cine-MRI frames were resampled with a voxel spacing of 1.5×1.5 ×3.15 mm and cropped to 128×128×32 around the center. The voxel intensity was normalized to the range from 0 to 1 through max-min normalization.

### 4.2. Implementation Details

Our method was implemented with PyTorch on an NVIDIA GeForce RTX 4090 GPU with 24 GB memory. We used an ADAM optimizer with a learning rate of 0.0001 and a batch size of 1. For brain image registration, our networks were trained for 100,000 iterations with inter-patient image pairs randomly picked from the training set. For cardiac image registration, our networks were first trained for 40,000 iterations with intra-patient image pairs that consist of two frames randomly picked from the same cine-MRI image. Then, the networks were trained for another 10,000 iterations with intra-patient image pairs consisting of only ED and ES frames, which optimizes the networks to register ED and ES frames. We performed validation after every 1,000 training iterations and used the model weights achieving the highest validation result for final testing. Our implementation code is publicly available at [https://github.com/MungoMeng/Registration-MAXCA](https://github.com/MungoMeng/Registration-MAXCA).

### 4.3. Experimental Designs

Our method was compared with fourteen existing deformable medical image registration methods, including traditional optimization-based registration methods and state-of-the-art deep registration methods. The included traditional methods are SyN (Avants et al. 2008) and NiftyReg (Modat et al. 2010), and we ran them using cross-correlation as the similarity measure. The included deep registration methods are VoxelMorph (Balakrishnan et al. 2019), TransMorph (Chen et al. 2022), Swin-VoxelMorph (Zhu and Lu 2022), TransMatch (Chen, Zheng, and Gee 2023), MLPMorph (Meng et al. 2023b), LapIRN (Mok and Chung 2020b), SDHNet (Zhou et al. 2023), ModeT (Wang, Ni, and Wang 2023), NICE-Net (Meng et al. 2022a), NICE-Trans (Meng et al. 2023a), Dual-PRNet++ (Kang et al. 2022), and CorrMLP (Meng et al. 2024). All deep registration methods were trained using the same loss functions as ours for a fair comparison.

We also conducted two ablation studies to further validate the effectiveness of our method. In the first ablation study, we replaced our MAXCA block with existing transformer or MLP blocks, including Swin transformer (Liu et al. 2021), Restormer (Zamir et al. 2022), Swin-MLP (Liu et al. 2021), Multi-axis gated MLP (Tu et al. 2022), Hire-MLP (Guo et al. 2022), and sMLP (Tang et al. 2022). In the second ablation study, we individually removed the global and local branches in our MAXCA block, and we also explored the contribution of the convolutional projection by replacing it with the common linear projection. The two ablation studies were performed with the direct registration architecture, and they also included a comparison baseline model that used Conv blocks at both the encoder and decoder.

Standard evaluation metrics for registration were adopted (Eisenmann et al. 2022; Hering et al. 2022): the registration accuracy was evaluated using the Dice similarity coefficients (DSC) of segmentation labels between the warped and fixed images, while the smoothness of spatial transformations was evaluated using the percentage of Negative Jacobian Determinants (NJD).

## 5. Results and Discussion
### 5.1. Comparison with Existing Methods

Table 1 presents the quantitative comparison with existing registration methods. The SA-based TransMorph and NICE-Trans achieved higher DSCs than the CNN-based VoxelMorph and NICE-Net, which validates the benefits of capturing long-range dependency for registration. Applying MLPs in MLPMorph and NICE-MLP further improved DSCs over SA-based registration methods. This is consistent with the previous findings (Meng et al. 2023b) showing that MLPs capture fine-grained long-range dependency at full resolution, enabling them to outperform existing SA-based registration methods. Nevertheless, our MAXCA block enabled higher DSCs than MLPs, producing consistent improvements in XCAMorph, NICE-XCA, and CorrXCA. By applying our MAXCA block in the state-of-the-art correlation-aware progressive registration architecture, our CorrXCA attained the highest DSCs among all the compared methods. We attribute the improvements to our optimized use of XCA, allowing for capturing both global and local long-range dependency among high-resolution features via attention mechanisms. Moreover, all deep registration methods achieved similar NJD results. This is because these methods adopted the same regularization settings, and this also implies that our method did not sacrifice transformation smoothness for registration accuracy.

Figure 3 presents the qualitative comparison for brain image registration, which shows that the warped images produced by our methods are more consistent with the fixed image. In addition, we report the runtime of each method in Appendix B, which shows that our XCAMorph, NICE-XCA, and CorrXCA are much faster than the traditional methods (SyN and NiftyReg) and only require similar runtime to the existing deep registration methods based on transformers or MLPs.

Table 1. Quantitative comparison for brain and cardiac image registration.

| Method | | Mindboggle (Brain) | | Buckner (Brain) | | ACDC (Cardiac) | |
|---|---|---|---|---|---|---|---|
| | | DSC ↑ | NJD (%)↓ | DSC ↑ | NJD (%)↓ | DSC ↑ | NJD (%)↓ |
| Before registration | | 0.347*,†,‡ | / | 0.406*,†,‡ | / | 0.590*,†,‡ | / |
| SyN | Traditional | 0.534*,†,‡ | 1.956 | 0.567*,†,‡ | 1.874 | 0.747*,†,‡ | 0.398 |
| NiftyReg | Traditional | 0.569*,†,‡ | 2.364 | 0.611*,†,‡ | 2.175 | 0.768*,†,‡ | 0.421 |
| VoxelMorph | CNN, Direct | 0.552* | 2.532 | 0.589* | 2.220 | 0.754* | 0.440 |
| TransMorph | SA, Direct | 0.571* | 2.400 | 0.608* | 2.183 | 0.768* | 0.492 |
| Swin-VoxelMorph | SA, Direct | 0.566* | 2.254 | 0.605* | 2.016 | 0.763* | 0.412 |
| TransMatch | SA, Direct | 0.578* | 2.036 | 0.622* | 1.995 | 0.770* | 0.425 |
| MLPMorph | MLP, Direct | 0.604* | 1.931 | 0.632* | 1.837 | 0.780* | 0.487 |
| **XCAMorph (Ours)** | XCA, Direct | 0.622 | **1.635** | 0.645 | **1.743** | 0.791 | **0.371** |
| LapIRN | CNN, Progressive | 0.605† | 2.164 | 0.632† | 2.112 | 0.790† | 0.454 |
| SDHNet | CNN, Progressive | 0.598† | 1.872 | 0.634† | 1.843 | 0.789† | 0.395 |
| NICE-Net | CNN, Progressive | 0.618† | 2.043 | 0.643† | 1.963 | 0.785† | 0.443 |
| ModeT | SA, Progressive | 0.613† | 1.849 | 0.642† | 1.843 | 0.790† | 0.395 |
| NICE-Trans | SA, Progressive | 0.625† | 2.324 | 0.649† | 2.277 | 0.799† | 0.473 |
| NICE-MLP | MLP, Progressive | 0.630† | 2.143 | 0.653† | 2.121 | 0.802† | 0.434 |
| **NICE-XCA (Ours)** | XCA, Progressive | <u>0.646</u> | <u>1.799</u> | <u>0.664</u> | 1.801 | <u>0.810</u> | 0.392 |
| Dual-PRNet++ | CNN, Correlation-aware | 0.608‡ | 2.424 | 0.636‡ | 2.195 | 0.777‡ | 0.479 |
| CorrMLP | MLP, Correlation-aware | 0.642‡ | 1.821 | 0.661‡ | 1.788 | <u>0.810</u>‡ | 0.389 |
| **CorrXCA (Ours)** | XCA, Correlation-aware | **0.655** | 1.815 | **0.671** | <u>1.762</u> | **0.817** | <u>0.382</u> |

**Bold**/<u>Underlined</u>: The best/second-best results in each column. ↑: the higher is better. ↓: the lower is better. */†/‡: *P*<0.05, in comparison to XCAMorph/NICE-XCA/CorrXCA.

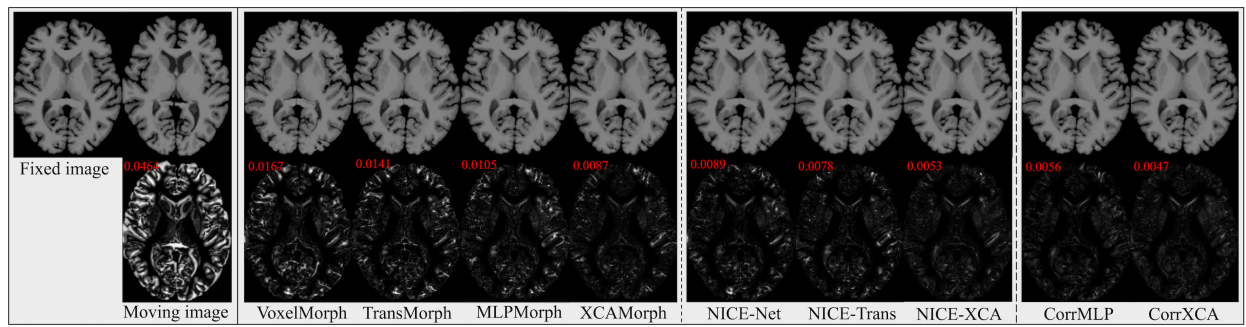

Figure 3. Qualitative comparison for brain image registration. Below each image is an error map that shows the intensity differences between the warped image and the fixed image, with the mean absolute error marked in red in the upper left corner. A cleaner error map indicates a better registration result.

## 5.2. Ablation Analysis on Network Blocks

Table 2 presents the DSC results of our ablation study on network blocks. The NJD results are omitted as all methods adopted the same regularization settings and achieved similar NJDs. All transformer and MLP-based blocks outperformed the CNN baseline, demonstrating the benefits of capturing long-range dependency for deformable image registration. Moreover, as Swin transformer blocks cannot be used at full resolution due to heavy loads of GPU memory (as reported in Appendix C), a Conv block was employed to process the full-resolution features, resulting in worse performance than other MLP/XCA-based blocks that can be used beginning from the full image resolution. This indicates the effectiveness of MLP/XCA in capturing long-range dependency at full resolution. The XCA-based Restormer failed to outperform MLP-based blocks, which is attributed to the fact that the Restormer was not designed for deformable medical image registration and lacks the capability to capture fine-grained local correspondence. By considering both global and local long-range dependency via our multi-axis design, our MAXCA block outperformed MLP-based blocks and achieved the highest results among all the compared network blocks.

Table 2. The DSC results of the ablation analysis on network blocks.

| Method | Mindboggle | Buckner | ACDC |
|---|---|---|---|
| CNN baseline | 0.556 | 0.590 | 0.755 |
| Swin transformer (SA) + Conv block | 0.585 | 0.616 | 0.770 |
| Restormer (XCA) | 0.595 | 0.624 | 0.774 |
| Swin-MLP | 0.603 | 0.632 | 0.778 |
| Multi-axis gated MLP | 0.604 | 0.632 | 0.780 |
| Hire-MLP | 0.599 | 0.627 | 0.775 |
| sMLP | 0.602 | 0.630 | 0.776 |
| **MAXCA (Ours)** | **0.622** | **0.645** | **0.791** |

**Bold**: The best result in each column is in bold.

## 5.3. Ablation Analysis within MAXCA

Table 3 presents the DSC results of our ablation study within the MAXCA block. The NJD results are omitted as all methods adopted the same regularization settings and achieved similar NJDs. Removing either global or local branch from our MAXCA block resulted in degraded performance, which shows the individual contribution of each branch and also suggests that both global and local long-range contexts are beneficial for medical image registration. Nevertheless, we found that local contexts are more crucial, as removing the local branch led to larger performance degradation. Further, compared to the commonly used linear projection, employing convolutional layers for the QKV projection resulted in better registration performance as it can enhance the local context before computing feature covariance.

Table 3. The DSC results of the ablation analysis within the MAXCA block.

| Method | Mindboggle | Buckner | ACDC |
|---|---|---|---|
| CNN baseline | 0.556 | 0.590 | 0.755 |
| w/o global branch | 0.608 | 0.636 | 0.783 |
| w/o local branch | 0.589 | 0.621 | 0.772 |
| w/ linear projection | 0.616 | 0.640 | 0.788 |
| **Ours** | **0.622** | **0.645** | **0.791** |

**Bold**: The best result in each column is in bold.

## 6. Conclusion

In this study, we have introduced a novel Multi-axis XCA (MAXCA) block to demonstrate the optimized use of XCA for deformable medical image registration. Our MAXCA block shows advantages in capturing both global and local long-range dependency among high-resolution image features via XCA, for finding dense pixel-wise correspondence between images. This block has been validated on three registration architectures and produced consistent improvements over state-of-the-art deformable image registration methods on the tasks of brain and cardiac image registration. We suggest that our MAXCA block can serve as a general network block applying to various network architectures for image registration tasks. Further validation could be performed with other registration network architectures, e.g., automatic fusion network (Meng et al. 2023c), or other registration tasks, e.g., brain tumor registration (Baheti et al. 2021; Meng et al. 2022c), in future studies.

## Acknowledgements

This work was supported by Australian Research Council (ARC) under Grant DP200103748.

## Appendix A: Architecture Settings

We report the architectural hyper-parameters used in the three exemplified registration network architectures in Table A1 (direct registration), Table A2 (progressive registration), and Table A3 (correlation-aware progressive registration). The architectural hyper-parameters include feature dimensions (channel numbers), head numbers of XCA, and region size $R$, at each pyramid scale.

The feature dimensions were chosen under the constraints of GPU memory (24 GB, NVIDIA GeForce RTX 4090 GPU), which could be increased to pursue better registration performance if larger GPU memory is available.

The XCA head numbers were chosen so that each attention head processed the features with 12 channels (e.g., the features with 24 channels were processed by two XCA heads). We also attempted to increase the XCA head numbers so that each attention head processed 8-channel features, but identified slightly degraded results.

The region size $R$ was chosen based on empirical experiments. We tried to set $R$ in a range from 4 to 12 and chose the $R$ that resulted in the highest validation results.

Table A1. Architectural hyper-parameters used in the direct registration architecture (XCAMorph).

| Hyper-parameter | Settings |
| --- | --- |
| Feature dimensions in the encoder | 24 – 48 – 96 – 192 – 384 |
| Feature dimensions in the decoder | 192 – 96 – 48 – 24 |
| XCA head numbers in the encoder | 2 – 4 – 8 – 16 – 32 |
| Region size $R$ in the encoder | 8 – 8 – 6 – 6 – 4 |

Table A2. Architectural hyper-parameters used in the progressive registration architecture (NICE-XCA).

| Hyper-parameter | Settings |
| --- | --- |
| Feature dimensions in the encoder | 12 – 24 – 48 – 96 |
| Feature dimensions in the decoder | 192 – 96 – 48 – 24 |
| XCA head numbers in the decoder | 16 – 8 – 4 – 2 |
| Region size $R$ in the decoder | 6 – 6 – 6 – 6 |

Table A3. Architectural hyper-parameters used in the correlation-aware progressive registration architecture (CorrXCA).

| Hyper-parameter | Settings |
|---|---|
| Feature dimensions in the encoder | 12 – 24 – 48 – 96 |
| Feature dimensions in the decoder | 192 – 96 – 48 – 24 |
| XCA head numbers in the decoder | 16 – 8 – 4 – 2 |
| Region size $R$ in the decoder | 6 – 6 – 6 – 6 |

## Appendix B: Registration Runtime

Table A4 presents the runtimes of the compared registration methods, where the inference time required to register a pair of images using GPU or CPU is reported. The GPU runtimes of traditional methods (SyN and NiftyReg) are unavailable for the lack of official GPU implementation.

As shown in Table A4, all deep registration methods are much faster than the traditional methods, allowing real-time registration with GPU (<0.5s for one image pair). Among deep registration methods, the methods based on transformers/MLP tend to be slower than the methods based on CNNs due to the additional computation required to model long-range dependency. The runtimes of our XCA-based methods (XCAMorph, NICE-XCA, and CorrMLP) are similar to their SA/MLP-based counterparts, suggesting that our method does not incur significant extra computational loads or registration runtime compared to the existing SA/MLP-based deep registration methods.

Table A4. Comparison of registration runtime (in seconds) for brain and cardiac image registration.

| Method | | Mindboggle/Buckner (Brain) | | ACDC (Cardiac) | |
|---|---|---|---|---|---|
| | | CPU (s) | GPU (s) | CPU (s) | GPU (s) |
| SyN | Traditional | 3427 | / | 401 | / |
| NiftyReg | Traditional | 159 | / | 112 | / |
| VoxelMorph | CNN, Direct | 2.84 | 0.23 | 0.36 | 0.02 |
| TransMorph | SA, Direct | 3.68 | 0.35 | 0.59 | 0.05 |
| Swin-VoxelMorph | SA, Direct | 5.67 | 0.52 | 0.91 | 0.08 |
| TransMatch | SA, Direct | 3.06 | 0.28 | 0.55 | 0.04 |
| MLPMorph | MLP, Direct | 4.21 | 0.37 | 0.61 | 0.05 |
| **XCAMorph (Ours)** | XCA, Direct | 4.42 | 0.40 | 0.62 | 0.05 |
| LapIRN | CNN, Progressive | 4.97 | 0.46 | 0.77 | 0.06 |
| SDHNet | CNN, Progressive | 3.24 | 0.26 | 0.45 | 0.03 |
| NICE-Net | CNN, Progressive | 3.55 | 0.32 | 0.49 | 0.04 |
| ModeT | SA, Progressive | 7.10 | 0.65 | 0.98 | 0.09 |
| NICE-Trans | SA, Progressive | 4.02 | 0.37 | 0.64 | 0.05 |
| NICE-MLP | MLP, Progressive | 3.81 | 0.34 | 0.57 | 0.05 |
| **NICE-XCA (Ours)** | XCA, Progressive | 3.95 | 0.38 | 0.59 | 0.05 |
| Dual-PRNet++ | CNN, Correlation-aware | 4.61 | 0.44 | 0.75 | 0.06 |
| CorrMLP | MLP, Correlation-aware | 5.48 | 0.49 | 0.83 | 0.07 |
| **CorrXCA (Ours)** | XCA, Correlation-aware | 5.54 | 0.50 | 0.84 | 0.07 |

# Appendix C: Memory Consumption

Table A5 shows the GPU memory consumption of the compared network blocks, where the GPU memory consumed during training for brain image registration is reported. The Swin transformer blocks were applied beginning from the half image resolution to reduce to GPU memory consumption, while the full-resolution features were processed by a Conv block. Even so, the hybrid use of Swin transformer and Conv blocks still used up the GPU memory (23.6GB out of 24GB), thus disabling the use of Swin transformer blocks to capture fine-grained long-range dependency among full-resolution anatomical details. In contrast, the MLP/XCA-based network blocks can be applied beginning from the full image resolution under the same constraints of GPU memory. Furthermore, compared to the existing MLP/XCA-based network blocks, our MAXCA block does not incur additional GPU memory consumption.

Table A5. Comparison of GPU memory consumption during training.

| Method | GPU Memory |
|---|---|
| Swin transformer (SA) + Conv block | 23.6GB |
| Restormer (XCA) | 23.2GB |
| Swin-MLP | 22.1GB |
| Multi-axis gated MLP | 22.8GB |
| Hire-MLP | 22.5GB |
| sMLP | 21.2GB |
| **MAXCA (Ours)** | 22.8GB |